\documentclass[english,10pt,aps,prd,a4paper,preprintnumbers,floatfix,nofootinbib,showpacs,superscriptaddress, notitlepage]{revtex4-1}
\usepackage[english]{babel}
\usepackage{amsmath}
\usepackage{graphicx}
\usepackage{color}
\usepackage{mathrsfs}   
\usepackage{amssymb}
\usepackage{hyperref}
\usepackage[normalem]{ulem} 
\usepackage{dcolumn}
\usepackage{bm}
\usepackage{graphicx}
\usepackage[sort&compress]{natbib}
\usepackage{epstopdf}
\newcommand {\be} {\begin{equation}}
\newcommand {\ee} {\end{equation}}

\definecolor{greenLinks}{rgb}{0, 0.6, 0} 
\definecolor{blueLinks}{rgb}{0, 0, 0.6}
\definecolor{redLinks}{rgb}{0.6, 0, 0}
\definecolor{tempText}{rgb}{0.55, 0.10,0.67}
\definecolor{eprintLinks}{rgb}{0.4, 0.4, 0.4}
\definecolor{journalLinks}{rgb}{0.6, 0, 0}
 
\def\vev#1{\left\langle #1\right\rangle}

\def\21{$\mathrm{SU(2)_L \otimes U(1)_Y}$ }
\def\31{$\mathrm{SU(3)_c \otimes U(1)_Q}$ }
\def\SM{$\mathrm{SU(3)_c \otimes SU(2)_L \otimes U(1)_Y}$ }

\def\3211{$\mathrm{SU(3) \otimes SU(2)_L \otimes U(1)_R \otimes U(1)_{B-L}}$ }
\def\321{$\mathrm{SU(3) \otimes SU(2) \otimes U(1)}$ }
\def\422{$\mathrm{SU(4) \otimes SU(2) \otimes SU(2)_R}$ }
\newcommand {\ignore}[1]{}

\newcommand{\sm}{{Standard Model }}

\def\vev#1{\left\langle #1\right\rangle}

\def\lnv{lepton number violation }

\newcommand{\AddrAHEP}{%
  AHEP Group, Institut de F\'{i}sica Corpuscular 
  (CSIC-Universitat de Val\`{e}ncia), Parc Cient\'ific de Paterna.\\
 C/ Catedr\'atico Jos\'e Beltr\'an, 2 E-46980 Paterna (Valencia) - SPAIN}

\begin{document}

\author{George Lazarides}\email{lazaride@eng.auth.gr}
\affiliation{School of Electrical and Computer Engineering, Faculty of Engineering,
Aristotle University of Thessaloniki, Thessaloniki 54124, Greece}
\author{Mario Reig}\email{mario.reig@ific.uv.es}
\affiliation{\AddrAHEP}
\author{Qaisar Shafi}\email{shafi@bartol.udel.edu}
\affiliation{Bartol Research Institute, Department of Physics and Astronomy,
University of Delaware, Newark, DE 19716, USA}
\affiliation{Institut f\"{u}r Theoretische Physik,
  Universit\"{a}t Heidelberg, Philosophenweg 16, Heidelberg, D-69120, Germany}
\author{Rahul Srivastava}\email{rahulsri@ific.uv.es}
\affiliation{\AddrAHEP}
\author{Jos\'{e} W. F. Valle}\email{valle@ific.uv.es}
\affiliation{\AddrAHEP}

\title{Spontaneous Breaking of Lepton Number and Cosmological Domain Wall Problem}


\begin{abstract}
\vspace{1cm}
 We show that if global lepton number symmetry is spontaneously broken in a post inflation epoch, then it can lead to the formation of cosmological domain walls. This happens in the well-known ``Majoron paradigm'' for neutrino mass generation.  We propose some realistic examples which allow spontaneous lepton number breaking to be safe from such domain walls.  
\end{abstract}

\maketitle

\medskip


\section{Introduction}
\label{sec:intro}


Topological defects such as monopoles, strings and domain walls~\cite{Zeldovich:1974uw, Vilenkin:1982ks} can arise in many gauge theories including grand unification. In addition, there can appear (hybrid) configurations such as monopoles connected via strings or walls bounded by strings. Two well known examples of the latter arise in $SO(10)$~\cite{Kibble:1982ae} and axion models~\cite{Kibble:1982dd}. Stable or sufficiently long lived domain walls, associated with symmetry breaking scales comparable to or larger than in the Standard Model (SM) will sooner or later become the dominant energy component of the early universe. As a consequence such domain walls pose a serious challenge in cosmology and should therefore be avoided in realistic model building (some possibilities were recently discussed in~\cite{Sato:2018nqy}). 

Domain walls are well known to appear associated with the spontaneous breaking of the Peccei-Quinn symmetry \cite{Sikivie:1982qv}. Here we note that also the ``weak'' $SU(2)_L$ may be associated with the presence of domain walls. This may happen in the context of spontaneous violation of lepton number symmetry. Indeed, such models in which lepton number is violated by a gauge singlet Higgs vacuum expectation value (vev) \cite{chikashige:1981ui,Schechter:1981cv} provides an attractive way to generate Majorana masses for neutrinos~\cite{Valle:2015pba}, as needed to account for current neutrino data~\cite{deSalas:2017kay}. 
In addition, it implies the existence of a physical Nambu-Goldstone boson, called Majoron. The latter may pick up a mass from explicit symmetry breaking by gravity effects~\cite{Rothstein:1992rh,Akhmedov:1992et,Akhmedov:1992hi,King:2017nbl}. Under such circumstances the Majoron may provide a good dark matter candidate~\cite{berezinsky:1993fm,Lattanzi:2007ux,Bazzocchi:2008fh,Lattanzi:2013uza,Lattanzi:2014mia,Kuo:2018fgw}~\footnote{Since gravitational effects are not calculable in a reliable way, here we prefer not to invoke their existence.}.

The origin of the domain wall problem in this case stems from the existence of an unbroken residual subgroup $Z_2$ arising from the spontaneous lepton number violation, which clashes with the unbroken $Z_3$ from the non-perturbative instanton effects associated with the weak $SU(2)_L$.
This implies that the domain wall problem associated with the weak $SU(2)_L$ exists in a broad class of Majoron models of neutrino mass generation. A standard mechanism for evading the domain wall problem is to invoke a suitable inflationary phase during their formation such that the walls are inflated away. In this letter we propose a more direct resolution of the domain wall problem which does not rely on inflation. We present various possible mechanisms for having realistic Majoron models, with and without supersymmetry, which allow spontaneous lepton number violation to occur without encountering a domain wall problem. 


\section{Global Lepton Number and Domain Wall problem}
\label{sec:ssb-of-lep-no}


Apart from the gauge symmetries it is well know that in the \sm there are two ``accidental'' global $U(1)$ symmetries namely the Baryon number $U(1)_B$ and the Lepton number $U(1)_L$ symmetries.
Although, accidental within \sm, these symmetries nonetheless play a very important role. 
The baryon number symmetry $U(1)_B$ is responsible for the stability of the proton and the lepton number symmetry plays a key role in neutrino mass generation and in determining the Dirac or Majorana nature of neutrinos.
Lepton number in the \sm is conserved at the Lagrangian level to all orders in perturbation theory. 
However, lepton number is an anomalous symmetry, hence it is explicitly broken by non-perturbative effects~\cite{tHooft:1976rip}
\footnote{Note that both baryon and lepton numbers are anomalous symmetries however a particular combination $U(1)_{B-L}$ is anomaly free. The other orthogonal combination $U(1)_{B+L}$ remains anomalous and hence it is explicitly broken by non-perturbative effects~\cite{tHooft:1976rip}.}.
In particular, owing to the $[SU(2)_L]^2\times U(1)_L$ anomaly, the non-perturbative instantons will explicitly break the initial lepton number symmetry $U(1)_L$ down to the discrete $Z_N$ subgroup, with
\begin{equation}
N = \sum_R N(R) \times L(R) \times T(R) = 3 \times 1 \times 1 = 3\,,
\label{eq:inst-z3}
\end{equation}
where $N(R)$ is the number of copies of a given fermion in representation R, $L(R)$ is the lepton number of the fermion and $T(R)$ is the $SU(2)_L$ Dynkin multiplicity index. 
For the $SU(2)_L$ group, the index $T(R)$ for the lowest representations, singlets, doublets and triplets are respectively, $T(1) = 0$,  $T(2) = 1$ and  $T(3) = 4$.

It is clear, from \eqref{eq:inst-z3}, that the non-perturbative instantons associated with the weak $SU(2)_L$ break $U(1)_L \to Z_3$.
Notice also that the threefold family replication in the \sm plays a crucial role in dictating the breaking $U(1)_L \to Z_3$. 
The residual $Z_3$ symmetry is exact at the classical and quantum level, implying the existence of degenerate vacua in our theory. 
Notice also that, in contrast to the case of axions, where the anomaly is related to the $U(1)$ Peccei-Quinn charge assignments, in the case of the lepton number there is an anomaly intrinsically associated with the chiral nature of weak $SU(2)_L$.

Although the tunneling rate from one vacuum to another due to instantons is extremely small at zero temperature\footnote{This rate is proportional to $exp(-2\pi/\alpha_W)$ and thus unimportant for our discussion \cite{Gross:1980br}}, sphaleron induced transitions between the vacua become relevant at higher temperatures \cite{Kuzmin:1985mm}. Moreover, it is argued that frequent transitions between the vacua occur even above the critical temperature $T_c$ for the electroweak transition\footnote{Our argument below involves only temperatures between around 200 GeV and $T_c$, where the sphalerons operate, and the transitions from one vacuum to another are very frequent \cite{Arnold:1987zg}.}. Thus, the B and L violating reactions at high temperatures are fast, so that the $U(1)_L$, $U(1)_B$ are explicitly broken by non-perturbative effects
down to discrete $Z_3$ symmetries. 

If the \sm is the final gauge theory, the non-perturbative breaking of lepton number won't be a serious issue. However, a dynamical understanding of the smallness of neutrinos mass often requires that lepton number is further broken down either explicitly or spontaneously by the new physics associated to neutrino mass generation.
A popular and well studied scenario is the case of spontaneous breaking of the lepton number~\cite{chikashige:1981ui,Schechter:1981cv}.
This is a specially attractive scenario that not only leads to Majorana masses, but also implies the existence of a Nambu-Goldstone boson, called Majoron. It breaks the global $U(1)_L$ lepton number symmetry down to a $Z_2$ subgroup through the vev of a \SM singlet scalar carrying two units of lepton number. 
However, we notice the mismatch between the unbroken residual subgroup $Z_2$ arising from the spontaneous lepton number violation and the subgroup $Z_3$ that is left unbroken by the nonperturbative effects.
Owing to this mismatch the domain walls will appear.

For temperatures between 200 GeV and the electroweak critical temperature
$T_c$ the tunneling rate between the vacua connected by the $Z_3$ subgroup which remains explicitly unbroken by instantons is very frequent.
The barrier separating different vacua, related by the $Z_3$, has static energy $E_{sph} (T)$, the sphaleron mass, and the width is of order $m_W^{-1}$. This is the size of the ``restricted instanton'' which minimizes the height of the barrier and corresponds to the sphaleron (see \cite{Lazarides:2001zd} and references therein). Since the wall thickness is much smaller than the horizon size at these temperatures, the walls are expected to be present. 
The mass per unit surface is $\simeq v^2m_W$, where $v$ is the order parameter, i.e. the vev that breaks $U(1)_L$. Even one such wall per horizon would provide an energy density $\simeq3v^2m_W/4t$ ($t$ is the cosmic time). This exceeds the radiation energy density $\rho_r=(\pi^2/30)g_*^{1/2}T^4$ ($g_*=106.75$ is the effective number of degrees of freedom) at a cosmic temperature $T>200$ GeV if the vev ($v$) is larger than about $(8\pi/3)^{1/2}(g_*/10)^{1/4}(m_W m_{\rm P})^{1/2} \simeq 8.2\times 10^{10}~{\rm GeV}$ ($m_{\rm P}=2.44\times 10^{18}~{\rm GeV}$ is the reduced Planck mass).
Such values for $v$ are very reasonable if they are to generate, say, the right handed neutrino masses within a type-I seesaw mechanism~\cite{Minkowski:1977sc,GellMann:1980vs,Yanagida:1979as,Schechter:1980gr,mohapatra:1980ia}.
Right after the wall domination, the Schwarzschild radius corresponding to the mass within each horizon becomes bigger than the horizon itself and the system becomes unstable and collapses into black holes leading to a cosmological catastrophe \cite{Vilenkin:1981zs}. 
%
%
Therefore, unless a suitable remedy is provided, we expect the standard high-scale type-I seesaw Majoron model of neutrino mass generation to be cosmologically inconsistent due to the existence of such domain walls. Note however that low-scale scenarios, such as the inverse seesaw Majoron schemes~\cite{Mohapatra:1986bd,GonzalezGarcia:1988rw,Joshipura:1992hp,Berezhiani:1992cd} constitute a potential way out. This is because in that case the lepton number violating order parameter can lie much below the electroweak 
scale, where sphaleron effects are negligible and $U(1)_L$ can be regarded as an exact continuous symmetry.\\[-.2cm]

The above spontaneous breaking of $U(1)_L \to Z_2$ by the vev of a field carrying two units of lepton number can be connected to neutrino masses in full generality at the operator level.
Consider the $U(1)_L$ invariant effective operator
\begin{equation}
\frac{1}{\Lambda^2} \bar{L}^c H H \sigma L\,.
\label{eq:maj-weinberg}
\end{equation}
In \eqref{eq:maj-weinberg} the field $L$ is the $SU(2)_L$ lepton doublet, $H$ is the Higgs doublet and $\sigma$ is a \sm gauge singlet
scalar field charged under the $U(1)_L$ symmetry.
Also, $\Lambda$ is the cutoff scale for the effective operator above which the full Ultra-Violet complete theory should be specified.
This operator is $U(1)_L$ invariant if $\sigma$ has charge $-2$ under the $U(1)_L$ symmetry. 
After $\sigma$ develops a non-zero vev, $\vev{\sigma}$, $U(1)_L$ is broken down to $Z_2$ and the expression in Eq. \eqref{eq:maj-weinberg} reduces to the famous Weinberg operator~\cite{weinberg:1980bf}. Again the CP odd part of $\sigma$ will be a Nambu-Goldstone boson, the Majoron. Here CP denotes the combined action of charge conjugation (C) and parity (P). 
%


\section{Solutions to the Domain Wall problem}
\label{sec:solutions}


In this section we consider alternative solutions to the domain wall problem which arises from the spontaneous breaking of $U(1)_L$ by the vev of a lepton-number-carrying scalar field. We focus attention on Majoron-type models characterized by the spontaneous breaking of lepton number at high-scale. 
The examples in Sec.\ref{sec:T}, \ref{sec:susyT} and \ref{sec:Diracon} involve only the Standard Model gauge structure. On the other hand the model considered in Sec.\ref{sec:family} requires an extension of the \sm with a gauge family symmetry.


\subsection{Majoron with Singlet-Triplet Seesaw}
\label{sec:T}

 The simplest solution of the domain wall problem in the Majoron model uses only the usual \sm gauge framework. It requires, in addition to the \sm fields, the following new ones with their \SM quantum numbers indicated in parenthesis and subscripts denoting their charges under $U(1)_L$:
 \begin{equation}
 \nu_R = (1,1,0)_{+1}\,, \,\, \Sigma_R = (1,3,0)_{+1}\,, \,\, \sigma = (1,1,0)_{-2}\,,
 \end{equation}
 where the first field ($\nu_R$) is a gauge singlet right-handed neutrino present in seesaw schemes~\cite{Minkowski:1977sc,GellMann:1980vs,Yanagida:1979as,Schechter:1980gr,mohapatra:1980ia} (with arbitrary multiplicity, which we take equal to one for simplicity, given that this is sufficient to account for the current neutrino oscillation data).
 The second field ($\Sigma_R$) is a $SU(2)_L$ triplet right-handed fermion
 and, the last field is the complex scalar whose vev $\vev{\sigma}$ is responsible for the spontaneous lepton number breaking. 
 The Lagrangian will now contain the following new couplings:
 \begin{equation}
 \mathcal{L}_{new}=y^{D i}_{\nu_R}\bar{L}^i \tilde{H} \nu_R \, + \,  y^{D i}_\Sigma \bar{L}^i \tilde{H} \Sigma_R 
 \, + \, y^M_\Sigma \sigma \bar{\Sigma}^c_R \Sigma_R \, + \, y^M_{\nu_R} \sigma \bar{\nu}^c_R \nu_R \,.
 \end{equation}
 where $\tilde{H} = i \tau_2 H^*$ with $\tau_2$ denoting the second Pauli matrix.
 After electroweak symmetry breaking the Higgs field will get a vev $\vev{ H} = v$ and we will have a seesaw-like mechanism for light neutrinos
 with mass matrix $m_\nu = M_D^T M_R^{-1} M_D$ where
 \begin{equation}
 \begin{split}
 &M_D=\left(\begin{array}{ccc}
 vy^{D1}_\nu&vy^{D2}_\nu&vy^{D3}_\nu\\
 vy^{D1}_\Sigma&vy^{D2}_\Sigma&vy^{D3}_\Sigma\\
 \end{array}\right)\,,\\&
 M_R=\left(\begin{array}{cc}
 y^M_{\nu_R}\vev{\sigma}&0\\
 0&y^M_\Sigma\vev{\sigma}\\
 \end{array}\right)\,.
 \end{split}
 \end{equation}
 
 The resulting matrix, $m_\nu$, has rank 2 leaving one light neutrino massless. Note that, since $\Sigma_R$ has non-trivial $SU(2)_L$ quantum numbers it produces a significant change in the $[SU(2)_L]^2\times U(1)_L$ anomaly, which is now given by
 \begin{equation}\label{anomaly}
N = \sum_R N(R) \times L(R) \times T(R) = 3\times 1\times 1-1\times 1\times 4 = -1~,
 \end{equation}
By computing the anomaly factor one sees that the  domain wall problem is absent in this extension. Therefore, the heavy triplet $\Sigma_R$ acts as an auxiliary Majorana field to address the domain wall issue. Moreover, it also acts as heavy messenger for small neutrino mass generation through the seesaw mechanism.


\subsection{ Majoron seesaw within supersymmetry}
\label{sec:susyT}

The simple solution illustrated in the previous section can be generalized within a supersymmetric (SUSY) context. We present here a simple supersymmetric model which also addresses the domain wall problem. The particle content and charges of the superfields are as shown in Table \ref{Tab:sec2}.
\begin{table}[ht]
\begin{center}
\begin{tabular}{| c c c c c c c | }
  \hline  
Superfields \hspace{0.5cm} & $SU(3)_C$  \hspace{0.5cm}  & $SU(2)_L$  \hspace{0.5cm}  &  $U(1)_Y$ \hspace{0.5cm} &  $U(1)_B$ \hspace{0.5cm} &	$U(1)_L$    \hspace{0.5cm} & $U(1)_R$   \hspace{0.5cm}     \\
\hline \hline
$Q_i$       \hspace{0.5cm} & 3	        \hspace{0.5cm}  & 2          \hspace{0.5cm}  &     1/6   \hspace{0.5cm} &    1/3    \hspace{0.5cm} &
0           \hspace{0.5cm} &  1         \hspace{0.5cm}                    \\
$u^c_i$     \hspace{0.5cm} & $\bar{3}$  \hspace{0.5cm}  & 1          \hspace{0.5cm}  &     -2/3  \hspace{0.5cm} &   -1/3    \hspace{0.5cm} &
0           \hspace{0.5cm} &  1         \hspace{0.5cm}                    \\
$d^c_i$     \hspace{0.5cm} & $\bar{3}$  \hspace{0.5cm}  & 1          \hspace{0.5cm}  &      1/3  \hspace{0.5cm} &   -1/3    \hspace{0.5cm} &
0           \hspace{0.5cm} &  1         \hspace{0.5cm}                    \\
\hline
$L_i$       \hspace{0.5cm} &  1         \hspace{0.5cm}  & 2          \hspace{0.5cm}  &     -1/2  \hspace{0.5cm} &      0    \hspace{0.5cm} &
1           \hspace{0.5cm} &  1         \hspace{0.5cm}                    \\
$e^c_i$     \hspace{0.5cm} &  1         \hspace{0.5cm}  & 1          \hspace{0.5cm}  &       1   \hspace{0.5cm} &      0    \hspace{0.5cm} &
-1          \hspace{0.5cm} &  1         \hspace{0.5cm}                    \\
$\nu^c_i$   \hspace{0.5cm} &  1         \hspace{0.5cm}  & 1          \hspace{0.5cm}  &       0   \hspace{0.5cm} &      0    \hspace{0.5cm} &
-1          \hspace{0.5cm} &  1         \hspace{0.5cm}                    \\
\hline
$T$         \hspace{0.5cm} &  1         \hspace{0.5cm}  & 3          \hspace{0.5cm}  &       1   \hspace{0.5cm} &      0    \hspace{0.5cm} &
-1          \hspace{0.5cm} &  1         \hspace{0.5cm}                    \\
$\bar{T}$   \hspace{0.5cm} &  1         \hspace{0.5cm}  & 3          \hspace{0.5cm}  &       -1  \hspace{0.5cm} &      0    \hspace{0.5cm} &
0           \hspace{0.5cm} &  1         \hspace{0.5cm}                    \\
\hline \hline
$H_u$       \hspace{0.5cm} &  1         \hspace{0.5cm}  & 2          \hspace{0.5cm}  &      1/2  \hspace{0.5cm} &      0    \hspace{0.5cm} &
0           \hspace{0.5cm} &  0         \hspace{0.5cm}                    \\
$H_d$       \hspace{0.5cm} &  1         \hspace{0.5cm}  & 2          \hspace{0.5cm}  &     -1/2  \hspace{0.5cm} &      0    \hspace{0.5cm} &
0           \hspace{0.5cm} &  0         \hspace{0.5cm}                    \\
\hline
$S$         \hspace{0.5cm} &  1         \hspace{0.5cm}  & 1          \hspace{0.5cm}  &       0   \hspace{0.5cm} &      0    \hspace{0.5cm} &
0           \hspace{0.5cm} &  2         \hspace{0.5cm}                    \\
$\phi$      \hspace{0.5cm} &  1         \hspace{0.5cm}  & 1          \hspace{0.5cm}  &       0   \hspace{0.5cm} &      0    \hspace{0.5cm} &
-1          \hspace{0.5cm} &  0         \hspace{0.5cm}                    \\
$\bar{\phi}$\hspace{0.5cm} &  1         \hspace{0.5cm}  & 1          \hspace{0.5cm}  &       0   \hspace{0.5cm} &      0    \hspace{0.5cm} &
1           \hspace{0.5cm} &  0         \hspace{0.5cm}                    
\\
  \hline
  \end{tabular}
\end{center}
\caption{ Particle content and charges. $\rm U(1)_R$ is an R-Symmetry under which the superpotential $\mathcal{W}$ has R-charge of $2$ units. }
 \label{Tab:sec2}
\end{table}

In addition to the usual minimal supersymmetric standard model (MSSM) superfields and the right-handed neutrinos ($\nu^c$), one adds the $SU(2)_L$ triplet superfields $T, \bar{T}$ and the gauge singlet superfields $S, \phi, \bar{\phi}$ with charges as listed in Table \ref{Tab:sec2}. 
The superpotential of our model is given by
\begin{eqnarray}
 \mathcal{W} & = &  \kappa \, S(\bar{\phi} \phi - M^2) \, + \, y^u_{ij} \, H_u Q_i u^c_j \, + \, y^d_{ij} \, H_d Q_i d^c_j  
 \, + \, y^\nu_{ij} \, H_u L_i \nu^c_j \, + \, y^e_{ij} \, H_d L_i e^c_j  \nonumber \\
& + & \lambda \, S H_u H_d  \, + \, y^T_i \, T L_i H_d \, + \,  y^{'T} \, \bar{\phi} T \bar{T} \, + \,   y^\phi_{ij} \, \frac{\bar{\phi}^2 \nu^c_i \nu^c_j}{m_P} \, ,
\label{spot}
\end{eqnarray}
where $i,j = 1,2,3$ are generation indices.

Owing to the presence of the triplet superfield $T$, the $[SU(2)_L]^2\times U(1)_L$ anomaly is again found to be
\begin{equation}\label{susy-anomaly}
N = \sum_R N(R) \times L(R) \times T(R) = 3\times 1\times 1-1\times 1\times 4 = -1\,.
 \end{equation}
Thus, unlike the usual Majoron models, here the instanton effects will break $U(1)_L \to Z_1$,  avoiding the domain wall problem. As in the previous case, this holds irrespective of the number of right-handed neutrino superfields, the minimal realistic model has just one.

Notice that this solution differs from the standard seesaw mechanism in that the Majoron coming from the imaginary parts of the $\phi,\bar\phi$ scalars carry one unit of lepton number, instead of two.
Moreover, our model has other attractive features which make it quite appealing. 
Apart from solving the domain wall problem, it automatically addresses the so-called ``$\mu$-problem'' of the MSSM~\cite{Dvali:1997uq}.
In addition we also have a R-symmetry which contains the usual R-parity of the MSSM, forbidding all the potentially dangerous terms in the superpotential, \eqref{spot}.
Finally, right-handed neutrino masses arise through the non-renormalizable term $\bar{\phi}^2 \nu^c \nu^c / m_P$, where we take the high scale as $m_P$.


\subsection{$SU(3)_{\rm lep}$ family symmetry for leptons}
\label{sec:family}

Consider now a $SU(3)_{\rm lep}$ gauge extension of the \sm scenario. Let quarks be singlets under this group, while leptons transform under it in a vector-like way\footnote{Note that this differs from the usual $SU(3)_{\rm lep}$ family symmetry used to address the observed fermion mass hierarchy \cite{Berezhiani:1983hm,Berezhiani:1990wn}.},
\begin{equation}
\begin{split}
& L = (1,2,-1/2,3)\,,\\&
e_R  = (1,1,-1,3)\,,\\&
\nu_R = (1,1,0,3)\,,
\end{split}
\end{equation}
with the first three entries in parenthesis indicating the standard model charges and the last entry the $SU(3)_{\rm lep}$ representation. This extension has several consequences. First of all, right-handed neutrinos cannot have a bare mass term. Their masses must be generated through the spontaneous violation of U(1)$_{\rm L}$. This is related with the breaking of  $SU(3)_{\rm lep}$ and is achieved by the vev of a flavour sextet scalar field $\sigma$ with lepton number $-2$ via the coupling
\begin{equation}
\sigma \bar{\nu}^c_R \nu_R\,.
\end{equation}

The second and more important implication is that this extension automatically solves the domain wall problem. The reason is that the center
of  $SU(3)_{\rm lep}$ which is  $Z_3$, exactly coincides with the discrete $Z_3$ subgroup of U(1)$_{\rm L}$ left unbroken by the anomaly. Since this accidental subgroup can be embedded in the continuous gauge group $SU(3)_{\rm lep}$, the degenerate minima are now connected by a gauge transformation, so that any difference among them becomes unphysical. In this way, the domain wall problem is solved. This is a Majoron variant of the domain wall axion solution given in the context of Grand Unified Theory (GUT) in Ref.~\cite{Lazarides:1982tw, Barr:1982bb}.


\subsection{Diracon Solution}
\label{sec:Diracon}


Another possible solution to the domain wall problem is obtained by enforcing that the spontaneous lepton number breaking is such that $U(1)_L \to Z_3$ instead of $ Z_2$. In this case there is no mismatch between the residual subgroup preserved by the anomaly and that preserved by the spontaneous \lnv due to $\vev{\sigma}$, so the domain wall problem will be automatically solved.
Clearly, the $U(1)_L \to Z_3$ spontaneous breaking cannot be accomplished within the framework of the canonical Majoron model.
In fact, if $Z_3$ is the residual unbroken symmetry then neutrinos cannot be Majorana particles. 
However, we note that for Dirac neutrinos the $U(1)_L \to Z_3$ breaking is viable, and will lead to a solution of the domain wall problem within a variant of the ``Diracon models''~\cite{Bonilla:2016zef,Bonilla:2016diq}.

To see this Diracon solution, the first thing is to realize that the lepton number of right handed neutrinos $\nu_R$ need not be the same as that of the left handed neutrinos~\cite{Bento:1991bc,Peltoniemi:1992ss}. 
In fact, a non-conventional lepton number assignment of $(4, 4, -5)$ for the three generations of $\nu_{i,R}$; $i = 1,2,3$, proposed in~\cite{Ma:2014qra,Ma:2015mjd} is equally acceptable. 

If the $\nu_{i,R}$ transform with such non-conventional charges under $U(1)_L$ then one cannot write down the tree level Dirac term 
$\bar{L} \tilde{H} \nu_{i,R}$ nor the Majoron Weinberg operator of \eqref{eq:maj-weinberg}.
However, one can still write down the following $U(1)_L$ invariant operators
\begin{equation}
\frac{1}{\Lambda} \bar{L} \tilde{H} \chi \nu_{i,R}\,, \quad \frac{1}{\Lambda^2} \bar{L} \tilde{H} \chi^* \chi^* \nu_{3,R} \, ,
\label{eq:dir-op}
\end{equation}
where $\nu_{i,R}$; $i = 1,2$ are the two right handed neutrinos carrying charge $4$ units under $U(1)_L$, and  $\nu_{3,R}$ has $U(1)_L$ charge of $-5$. Also, the field $\chi$ has charge of $-3$ under $U(1)_L$.
It can be easily seen that the vev of the $\chi$ field will spontaneously break $U(1)_L \to Z_3$ with the resulting neutrinos being Dirac in nature. 
Furthermore, the CP odd part of $\chi$ will be a Nambu-Goldstone boson which we call Diracon and is associated with the Dirac mass generation of the neutrinos.
Now, since the $U(1)_L$ in this case is spontaneously broken to the same residual subgroup $Z_3$ as that preserved by the non-perturbative $SU(2)_L$ instantons, there is no mismatch and hence the problem of domain walls is automatically avoided.


\section{Conclusions}
\label{sec:con}


 We have shown that if the global lepton number symmetry is broken spontaneously in a post-inflationary epoch, then it can lead to the formation of cosmological domain walls. Since the presence  of these domain walls may spoil the standard picture of cosmological evolution, we have studied the conditions to prevent their formation as a result of spontaneous symmetry breaking. We have shown that the simplest seesaw Majoron models of neutrino masses have, in principle, a domain wall problem associated with the chiral $SU(2)_L$ gauge group describing the weak interaction.  We have also provided some explicit and realistic solutions which allow a safe spontaneous breaking of lepton number, free of domain walls. Some of these models involve new particles that could potentially lead to phenomenological implications. 

\begin{acknowledgments}

The work of M.R., R.S. and J.W.F.V is supported by the Spanish grants FPA2017-85216-P (AEI/FEDER, UE), SEV-2014-0398 and  PROMETEO/2018/165 (Generalitat  Valenciana) and the Spanish Red Consolider MultiDark FPA2017-90566-REDC. M.R. is also supported by Formaci\'on de Profesorado Universitario (FPU) grant FPU16/01907. 
Q.S. is supported in part by the DOE grant, No. DE-SC0013880. He also thanks the Alexander von Humboldt Foundation for providing support
through their research prize. Q.S. also acknowledges hospitality of the Instit\"{u}t f\"{u}r Theoretische Physik during his stay in Heidelberg.

\end{acknowledgments}
%

\begin{thebibliography}{47}%
	\makeatletter
	\providecommand \@ifxundefined [1]{%
		\@ifx{#1\undefined}
	}%
	\providecommand \@ifnum [1]{%
		\ifnum #1\expandafter \@firstoftwo
		\else \expandafter \@secondoftwo
		\fi
	}%
	\providecommand \@ifx [1]{%
		\ifx #1\expandafter \@firstoftwo
		\else \expandafter \@secondoftwo
		\fi
	}%
	\providecommand \natexlab [1]{#1}%
	\providecommand \enquote  [1]{``#1''}%
	\providecommand \bibnamefont  [1]{#1}%
	\providecommand \bibfnamefont [1]{#1}%
	\providecommand \citenamefont [1]{#1}%
	\providecommand \href@noop [0]{\@secondoftwo}%
	\providecommand \href [0]{\begingroup \@sanitize@url \@href}%
	\providecommand \@href[1]{\@@startlink{#1}\@@href}%
	\providecommand \@@href[1]{\endgroup#1\@@endlink}%
	\providecommand \@sanitize@url [0]{\catcode `\\12\catcode `\$12\catcode
		`\&12\catcode `\#12\catcode `\^12\catcode `\_12\catcode `\%12\relax}%
	\providecommand \@@startlink[1]{}%
	\providecommand \@@endlink[0]{}%
	\providecommand \url  [0]{\begingroup\@sanitize@url \@url }%
	\providecommand \@url [1]{\endgroup\@href {#1}{\urlprefix }}%
	\providecommand \urlprefix  [0]{URL }%
	\providecommand \Eprint [0]{\href }%
	\providecommand \doibase [0]{http://dx.doi.org/}%
	\providecommand \selectlanguage [0]{\@gobble}%
	\providecommand \bibinfo  [0]{\@secondoftwo}%
	\providecommand \bibfield  [0]{\@secondoftwo}%
	\providecommand \translation [1]{[#1]}%
	\providecommand \BibitemOpen [0]{}%
	\providecommand \bibitemStop [0]{}%
	\providecommand \bibitemNoStop [0]{.\EOS\space}%
	\providecommand \EOS [0]{\spacefactor3000\relax}%
	\providecommand \BibitemShut  [1]{\csname bibitem#1\endcsname}%
	\let\auto@bib@innerbib\@empty
	\bibitem [{\citenamefont {Zeldovich}\ \emph {et~al.}(1974)\citenamefont
		{Zeldovich}, \citenamefont {Kobzarev},\ and\ \citenamefont
		{Okun}}]{Zeldovich:1974uw}%
	\BibitemOpen
	\bibfield  {author} {\bibinfo {author} {\bibfnamefont {{\relax Ya}.~B.}\
			\bibnamefont {Zeldovich}}, \bibinfo {author} {\bibfnamefont {I.~{\relax
					Yu}.}\ \bibnamefont {Kobzarev}}, \ and\ \bibinfo {author} {\bibfnamefont
			{L.~B.}\ \bibnamefont {Okun}},\ }\href@noop {} {\bibfield  {journal}
		{\bibinfo  {journal} {Zh. Eksp. Teor. Fiz.}\ }\textbf {\bibinfo {volume}
			{67}},\ \bibinfo {pages} {3} (\bibinfo {year} {1974})},\ \bibinfo {note}
	{[Sov. Phys. JETP40,1(1974)]}\BibitemShut {NoStop}%
	\bibitem [{\citenamefont {Vilenkin}\ and\ \citenamefont
		{Everett}(1982)}]{Vilenkin:1982ks}%
	\BibitemOpen
	\bibfield  {author} {\bibinfo {author} {\bibfnamefont {A.}~\bibnamefont
			{Vilenkin}}\ and\ \bibinfo {author} {\bibfnamefont {A.~E.}\ \bibnamefont
			{Everett}},\ }\href {\doibase 10.1103/PhysRevLett.48.1867} {\bibfield
		{journal} {\bibinfo  {journal} {Phys. Rev. Lett.}\ }\textbf {\bibinfo
			{volume} {48}},\ \bibinfo {pages} {1867} (\bibinfo {year}
		{1982})}\BibitemShut {NoStop}%
	\bibitem [{\citenamefont {Kibble}\ \emph
		{et~al.}(1982{\natexlab{a}})\citenamefont {Kibble}, \citenamefont
		{Lazarides},\ and\ \citenamefont {Shafi}}]{Kibble:1982ae}%
	\BibitemOpen
	\bibfield  {author} {\bibinfo {author} {\bibfnamefont {T.~W.~B.}\
			\bibnamefont {Kibble}}, \bibinfo {author} {\bibfnamefont {G.}~\bibnamefont
			{Lazarides}}, \ and\ \bibinfo {author} {\bibfnamefont {Q.}~\bibnamefont
			{Shafi}},\ }\href {\doibase 10.1016/0370-2693(82)90829-2} {\bibfield
		{journal} {\bibinfo  {journal} {Phys. Lett.}\ }\textbf {\bibinfo {volume}
			{113B}},\ \bibinfo {pages} {237} (\bibinfo {year}
		{1982}{\natexlab{a}})}\BibitemShut {NoStop}%
	\bibitem [{\citenamefont {Kibble}\ \emph
		{et~al.}(1982{\natexlab{b}})\citenamefont {Kibble}, \citenamefont
		{Lazarides},\ and\ \citenamefont {Shafi}}]{Kibble:1982dd}%
	\BibitemOpen
	\bibfield  {author} {\bibinfo {author} {\bibfnamefont {T.~W.~B.}\
			\bibnamefont {Kibble}}, \bibinfo {author} {\bibfnamefont {G.}~\bibnamefont
			{Lazarides}}, \ and\ \bibinfo {author} {\bibfnamefont {Q.}~\bibnamefont
			{Shafi}},\ }\href {\doibase 10.1103/PhysRevD.26.435} {\bibfield  {journal}
		{\bibinfo  {journal} {Phys. Rev.}\ }\textbf {\bibinfo {volume} {D26}},\
		\bibinfo {pages} {435} (\bibinfo {year} {1982}{\natexlab{b}})}\BibitemShut
	{NoStop}%
	\bibitem [{\citenamefont {Sato}\ \emph {et~al.}(2018)\citenamefont {Sato},
		\citenamefont {Takahashi},\ and\ \citenamefont {Yamada}}]{Sato:2018nqy}%
	\BibitemOpen
	\bibfield  {author} {\bibinfo {author} {\bibfnamefont {R.}~\bibnamefont
			{Sato}}, \bibinfo {author} {\bibfnamefont {F.}~\bibnamefont {Takahashi}}, \
		and\ \bibinfo {author} {\bibfnamefont {M.}~\bibnamefont {Yamada}},\
	}\href@noop {} {\  (\bibinfo {year} {2018})},\ \Eprint
	{http://arxiv.org/abs/1805.10533} {arXiv:1805.10533 [hep-ph]} \BibitemShut
	{NoStop}%
	\bibitem [{\citenamefont {Sikivie}(1982)}]{Sikivie:1982qv}%
	\BibitemOpen
	\bibfield  {author} {\bibinfo {author} {\bibfnamefont {P.}~\bibnamefont
			{Sikivie}},\ }\href {\doibase 10.1103/PhysRevLett.48.1156} {\bibfield
		{journal} {\bibinfo  {journal} {Phys. Rev. Lett.}\ }\textbf {\bibinfo
			{volume} {48}},\ \bibinfo {pages} {1156} (\bibinfo {year}
		{1982})}\BibitemShut {NoStop}%
	\bibitem [{\citenamefont {Chikashige}\ \emph {et~al.}(1981)\citenamefont
		{Chikashige}, \citenamefont {Mohapatra},\ and\ \citenamefont
		{Peccei}}]{chikashige:1981ui}%
	\BibitemOpen
	\bibfield  {author} {\bibinfo {author} {\bibfnamefont {Y.}~\bibnamefont
			{Chikashige}}, \bibinfo {author} {\bibfnamefont {R.~N.}\ \bibnamefont
			{Mohapatra}}, \ and\ \bibinfo {author} {\bibfnamefont {R.~D.}\ \bibnamefont
			{Peccei}},\ }\href@noop {} {\bibfield  {journal} {\bibinfo  {journal} {Phys.
				Lett.}\ }\textbf {\bibinfo {volume} {B98}},\ \bibinfo {pages} {265} (\bibinfo
		{year} {1981})}\BibitemShut {NoStop}%
	\bibitem [{\citenamefont {Schechter}\ and\ \citenamefont
		{Valle}(1982)}]{Schechter:1981cv}%
	\BibitemOpen
	\bibfield  {author} {\bibinfo {author} {\bibfnamefont {J.}~\bibnamefont
			{Schechter}}\ and\ \bibinfo {author} {\bibfnamefont {J.~W.~F.}\ \bibnamefont
			{Valle}},\ }\href {\doibase 10.1103/PhysRevD.25.774} {\bibfield  {journal}
		{\bibinfo  {journal} {Phys. Rev.}\ }\textbf {\bibinfo {volume} {D25}},\
		\bibinfo {pages} {774} (\bibinfo {year} {1982})}\BibitemShut {NoStop}%
	\bibitem [{\citenamefont {Valle}\ and\ \citenamefont
		{Romao}()}]{Valle:2015pba}%
	\BibitemOpen
	\bibfield  {author} {\bibinfo {author} {\bibfnamefont {J.~W.~F.}\
			\bibnamefont {Valle}}\ and\ \bibinfo {author} {\bibfnamefont {J.~C.}\
			\bibnamefont {Romao}},\ }\href@noop {} {\emph {\bibinfo {title} {{Neutrinos
					in high energy and astroparticle physics}}}}\ (\bibinfo  {publisher} {John
		Wiley \& Sons (2015),~ note =})\BibitemShut {NoStop}%
	\bibitem [{\citenamefont {de~Salas}\ \emph {et~al.}(2018)\citenamefont
		{de~Salas} \emph {et~al.}}]{deSalas:2017kay}%
	\BibitemOpen
	\bibfield  {author} {\bibinfo {author} {\bibfnamefont {P.~F.}\ \bibnamefont
			{de~Salas}} \emph {et~al.},\ }\href {\doibase 10.1016/j.physletb.2018.06.019}
	{\bibfield  {journal} {\bibinfo  {journal} {Phys. Lett.}\ }\textbf {\bibinfo
			{volume} {B782}},\ \bibinfo {pages} {633} (\bibinfo {year} {2018})},\
	\bibinfo {note} {\url{http://globalfit.astroparticles.es/}},\ \Eprint
	{http://arxiv.org/abs/1708.01186} {arXiv:1708.01186 [hep-ph]} \BibitemShut
	{NoStop}%
	\bibitem [{\citenamefont {Rothstein}\ \emph {et~al.}(1993)\citenamefont
		{Rothstein}, \citenamefont {Babu},\ and\ \citenamefont
		{Seckel}}]{Rothstein:1992rh}%
	\BibitemOpen
	\bibfield  {author} {\bibinfo {author} {\bibfnamefont {I.~Z.}\ \bibnamefont
			{Rothstein}}, \bibinfo {author} {\bibfnamefont {K.~S.}\ \bibnamefont {Babu}},
		\ and\ \bibinfo {author} {\bibfnamefont {D.}~\bibnamefont {Seckel}},\ }\href
	{\doibase 10.1016/0550-3213(93)90368-Y} {\bibfield  {journal} {\bibinfo
			{journal} {Nucl. Phys.}\ }\textbf {\bibinfo {volume} {B403}},\ \bibinfo
		{pages} {725} (\bibinfo {year} {1993})},\ \Eprint
	{http://arxiv.org/abs/hep-ph/9301213} {arXiv:hep-ph/9301213 [hep-ph]}
	\BibitemShut {NoStop}%
	\bibitem [{\citenamefont {Akhmedov}\ \emph
		{et~al.}(1993{\natexlab{a}})\citenamefont {Akhmedov}, \citenamefont
		{Berezhiani}, \citenamefont {Senjanovic},\ and\ \citenamefont
		{Tao}}]{Akhmedov:1992et}%
	\BibitemOpen
	\bibfield  {author} {\bibinfo {author} {\bibfnamefont {E.~K.}\ \bibnamefont
			{Akhmedov}}, \bibinfo {author} {\bibfnamefont {Z.~G.}\ \bibnamefont
			{Berezhiani}}, \bibinfo {author} {\bibfnamefont {G.}~\bibnamefont
			{Senjanovic}}, \ and\ \bibinfo {author} {\bibfnamefont {Z.-j.}\ \bibnamefont
			{Tao}},\ }\href {\doibase 10.1103/PhysRevD.47.3245} {\bibfield  {journal}
		{\bibinfo  {journal} {Phys. Rev.}\ }\textbf {\bibinfo {volume} {D47}},\
		\bibinfo {pages} {3245} (\bibinfo {year} {1993}{\natexlab{a}})},\ \Eprint
	{http://arxiv.org/abs/hep-ph/9208230} {arXiv:hep-ph/9208230 [hep-ph]}
	\BibitemShut {NoStop}%
	\bibitem [{\citenamefont {Akhmedov}\ \emph
		{et~al.}(1993{\natexlab{b}})\citenamefont {Akhmedov}, \citenamefont
		{Berezhiani}, \citenamefont {Mohapatra},\ and\ \citenamefont
		{Senjanovic}}]{Akhmedov:1992hi}%
	\BibitemOpen
	\bibfield  {author} {\bibinfo {author} {\bibfnamefont {E.~K.}\ \bibnamefont
			{Akhmedov}}, \bibinfo {author} {\bibfnamefont {Z.~G.}\ \bibnamefont
			{Berezhiani}}, \bibinfo {author} {\bibfnamefont {R.~N.}\ \bibnamefont
			{Mohapatra}}, \ and\ \bibinfo {author} {\bibfnamefont {G.}~\bibnamefont
			{Senjanovic}},\ }\href {\doibase 10.1016/0370-2693(93)90887-N} {\bibfield
		{journal} {\bibinfo  {journal} {Phys. Lett.}\ }\textbf {\bibinfo {volume}
			{B299}},\ \bibinfo {pages} {90} (\bibinfo {year} {1993}{\natexlab{b}})},\
	\Eprint {http://arxiv.org/abs/hep-ph/9209285} {arXiv:hep-ph/9209285 [hep-ph]}
	\BibitemShut {NoStop}%
	\bibitem [{\citenamefont {King}\ and\ \citenamefont
		{Ludl}(2017)}]{King:2017nbl}%
	\BibitemOpen
	\bibfield  {author} {\bibinfo {author} {\bibfnamefont {S.~F.}\ \bibnamefont
			{King}}\ and\ \bibinfo {author} {\bibfnamefont {P.~O.}\ \bibnamefont
			{Ludl}},\ }\href {\doibase 10.1007/JHEP03(2017)174} {\bibfield  {journal}
		{\bibinfo  {journal} {JHEP}\ }\textbf {\bibinfo {volume} {03}},\ \bibinfo
		{pages} {174} (\bibinfo {year} {2017})},\ \Eprint
	{http://arxiv.org/abs/1701.04794} {arXiv:1701.04794 [hep-ph]} \BibitemShut
	{NoStop}%
	\bibitem [{\citenamefont {Berezinsky}\ and\ \citenamefont
		{Valle}()}]{berezinsky:1993fm}%
	\BibitemOpen
	\bibfield  {author} {\bibinfo {author} {\bibfnamefont {V.}~\bibnamefont
			{Berezinsky}}\ and\ \bibinfo {author} {\bibfnamefont {J.~W.~F.}\ \bibnamefont
			{Valle}},\ }\href@noop {} {\bibfield  {journal} {\bibinfo  {journal} {Phys.
				Lett.}\ }\textbf {\bibinfo {volume} {B318}},\ \bibinfo {pages} {360}},\
	\Eprint {http://arxiv.org/abs/hep-ph/9309214} {hep-ph/9309214} \BibitemShut
	{NoStop}%
	\bibitem [{\citenamefont {Lattanzi}\ and\ \citenamefont
		{Valle}()}]{Lattanzi:2007ux}%
	\BibitemOpen
	\bibfield  {author} {\bibinfo {author} {\bibfnamefont {M.}~\bibnamefont
			{Lattanzi}}\ and\ \bibinfo {author} {\bibfnamefont {J.~W.~F.}\ \bibnamefont
			{Valle}},\ }\href@noop {} {\bibfield  {journal} {\bibinfo  {journal} {Phys.
				Rev. Lett.}\ }\textbf {\bibinfo {volume} {99}},\ \bibinfo {pages} {121301}},\
	\Eprint {http://arxiv.org/abs/arXiv:0705.2406 [astro-ph]} {arXiv:0705.2406
		[astro-ph]} \BibitemShut {NoStop}%
	\bibitem [{\citenamefont {Bazzocchi}\ \emph {et~al.}(2008)\citenamefont
		{Bazzocchi} \emph {et~al.}}]{Bazzocchi:2008fh}%
	\BibitemOpen
	\bibfield  {author} {\bibinfo {author} {\bibfnamefont {F.}~\bibnamefont
			{Bazzocchi}} \emph {et~al.},\ }\href {\doibase 10.1088/1475-7516/2008/08/013}
	{\bibfield  {journal} {\bibinfo  {journal} {JCAP}\ }\textbf {\bibinfo
			{volume} {0808}},\ \bibinfo {pages} {013} (\bibinfo {year} {2008})},\ \Eprint
	{http://arxiv.org/abs/0805.2372} {arXiv:0805.2372 [astro-ph]} \BibitemShut
	{NoStop}%
	\bibitem [{\citenamefont {Lattanzi}\ \emph {et~al.}()\citenamefont {Lattanzi}
		\emph {et~al.}}]{Lattanzi:2013uza}%
	\BibitemOpen
	\bibfield  {author} {\bibinfo {author} {\bibfnamefont {M.}~\bibnamefont
			{Lattanzi}} \emph {et~al.},\ }\href {\doibase 10.1103/PhysRevD.88.063528}
	{\bibfield  {journal} {\bibinfo  {journal} {Phys.Rev.}\ }\textbf {\bibinfo
			{volume} {D88}},\ \bibinfo {pages} {063528}},\ \Eprint
	{http://arxiv.org/abs/1303.4685} {arXiv:1303.4685 [astro-ph.HE]} \BibitemShut
	{NoStop}%
	\bibitem [{\citenamefont {Lattanzi}\ \emph {et~al.}(2014)\citenamefont
		{Lattanzi}, \citenamefont {Lineros},\ and\ \citenamefont
		{Taoso}}]{Lattanzi:2014mia}%
	\BibitemOpen
	\bibfield  {author} {\bibinfo {author} {\bibfnamefont {M.}~\bibnamefont
			{Lattanzi}}, \bibinfo {author} {\bibfnamefont {R.~A.}\ \bibnamefont
			{Lineros}}, \ and\ \bibinfo {author} {\bibfnamefont {M.}~\bibnamefont
			{Taoso}},\ }\href {\doibase 10.1088/1367-2630/16/12/125012} {\bibfield
		{journal} {\bibinfo  {journal} {New J. Phys.}\ }\textbf {\bibinfo {volume}
			{16}},\ \bibinfo {pages} {125012} (\bibinfo {year} {2014})},\ \Eprint
	{http://arxiv.org/abs/1406.0004} {arXiv:1406.0004 [hep-ph]} \BibitemShut
	{NoStop}%
	\bibitem [{\citenamefont {Kuo}\ \emph {et~al.}(2018)\citenamefont {Kuo} \emph
		{et~al.}}]{Kuo:2018fgw}%
	\BibitemOpen
	\bibfield  {author} {\bibinfo {author} {\bibfnamefont {J.-L.}\ \bibnamefont
			{Kuo}} \emph {et~al.},\ }\href@noop {} {\  (\bibinfo {year} {2018})},\
	\Eprint {http://arxiv.org/abs/1803.05650} {arXiv:1803.05650 [astro-ph.CO]}
	\BibitemShut {NoStop}%
	\bibitem [{\citenamefont {'t~Hooft}(1976)}]{tHooft:1976rip}%
	\BibitemOpen
	\bibfield  {author} {\bibinfo {author} {\bibfnamefont {G.}~\bibnamefont
			{'t~Hooft}},\ }\href {\doibase 10.1103/PhysRevLett.37.8} {\bibfield
		{journal} {\bibinfo  {journal} {Phys. Rev. Lett.}\ }\textbf {\bibinfo
			{volume} {37}},\ \bibinfo {pages} {8} (\bibinfo {year} {1976})}\BibitemShut
	{NoStop}%
	\bibitem [{\citenamefont {Gross}\ \emph {et~al.}(1981)\citenamefont {Gross},
		\citenamefont {Pisarski},\ and\ \citenamefont {Yaffe}}]{Gross:1980br}%
	\BibitemOpen
	\bibfield  {author} {\bibinfo {author} {\bibfnamefont {D.~J.}\ \bibnamefont
			{Gross}}, \bibinfo {author} {\bibfnamefont {R.~D.}\ \bibnamefont {Pisarski}},
		\ and\ \bibinfo {author} {\bibfnamefont {L.~G.}\ \bibnamefont {Yaffe}},\
	}\href {\doibase 10.1103/RevModPhys.53.43} {\bibfield  {journal} {\bibinfo
		{journal} {Rev. Mod. Phys.}\ }\textbf {\bibinfo {volume} {53}},\ \bibinfo
	{pages} {43} (\bibinfo {year} {1981})}\BibitemShut {NoStop}%
\bibitem [{\citenamefont {Kuzmin}\ \emph {et~al.}(1985)\citenamefont {Kuzmin},
	\citenamefont {Rubakov},\ and\ \citenamefont {Shaposhnikov}}]{Kuzmin:1985mm}%
\BibitemOpen
\bibfield  {author} {\bibinfo {author} {\bibfnamefont {V.~A.}\ \bibnamefont
		{Kuzmin}}, \bibinfo {author} {\bibfnamefont {V.~A.}\ \bibnamefont {Rubakov}},
	\ and\ \bibinfo {author} {\bibfnamefont {M.~E.}\ \bibnamefont
		{Shaposhnikov}},\ }\href {\doibase 10.1016/0370-2693(85)91028-7} {\bibfield
	{journal} {\bibinfo  {journal} {Phys. Lett.}\ }\textbf {\bibinfo {volume}
		{155B}},\ \bibinfo {pages} {36} (\bibinfo {year} {1985})}\BibitemShut
{NoStop}%
\bibitem [{\citenamefont {Arnold}\ and\ \citenamefont
	{McLerran}(1988)}]{Arnold:1987zg}%
\BibitemOpen
\bibfield  {author} {\bibinfo {author} {\bibfnamefont {P.~B.}\ \bibnamefont
		{Arnold}}\ and\ \bibinfo {author} {\bibfnamefont {L.~D.}\ \bibnamefont
		{McLerran}},\ }\href {\doibase 10.1103/PhysRevD.37.1020} {\bibfield
	{journal} {\bibinfo  {journal} {Phys. Rev.}\ }\textbf {\bibinfo {volume}
		{D37}},\ \bibinfo {pages} {1020} (\bibinfo {year} {1988})}\BibitemShut
{NoStop}%
\bibitem [{\citenamefont {Lazarides}(2002)}]{Lazarides:2001zd}%
\BibitemOpen
\bibfield  {author} {\bibinfo {author} {\bibfnamefont {G.}~\bibnamefont
		{Lazarides}},\ }\bibfield  {booktitle} {\emph {\bibinfo {booktitle}
		{{Cosmological crossroads: An advanced course in mathematical, physical and
				string cosmology. Proceedings, 1st Aegean Summer School, Karlovasi, Samos,
				Greece, September 21-29, 2001}}},\ }\href@noop {} {\bibfield  {journal}
	{\bibinfo  {journal} {Lect. Notes Phys.}\ }\textbf {\bibinfo {volume}
		{592}},\ \bibinfo {pages} {351} (\bibinfo {year} {2002})},\ \Eprint
{http://arxiv.org/abs/hep-ph/0111328} {arXiv:hep-ph/0111328 [hep-ph]}
\BibitemShut {NoStop}%
\bibitem [{\citenamefont {Minkowski}(1977)}]{Minkowski:1977sc}%
\BibitemOpen
\bibfield  {author} {\bibinfo {author} {\bibfnamefont {P.}~\bibnamefont
		{Minkowski}},\ }\href@noop {} {\bibfield  {journal} {\bibinfo  {journal}
		{Phys. Lett.}\ }\textbf {\bibinfo {volume} {B67}},\ \bibinfo {pages} {421}
	(\bibinfo {year} {1977})}\BibitemShut {NoStop}%
\bibitem [{\citenamefont {Gell-Mann}\ \emph {et~al.}(1979)\citenamefont
	{Gell-Mann}, \citenamefont {Ramond},\ and\ \citenamefont
	{Slansky}}]{GellMann:1980vs}%
\BibitemOpen
\bibfield  {author} {\bibinfo {author} {\bibfnamefont {M.}~\bibnamefont
		{Gell-Mann}}, \bibinfo {author} {\bibfnamefont {P.}~\bibnamefont {Ramond}}, \
	and\ \bibinfo {author} {\bibfnamefont {R.}~\bibnamefont {Slansky}},\
}\bibfield  {booktitle} {\emph {\bibinfo {booktitle} {Supergravity Workshop
		Stony Brook, New York, September 27-28, 1979}},\ }\href@noop {} {\bibfield
{journal} {\bibinfo  {journal} {Conf. Proc.}\ }\textbf {\bibinfo {volume}
	{C790927}},\ \bibinfo {pages} {315} (\bibinfo {year} {1979})},\ \Eprint
{http://arxiv.org/abs/1306.4669} {arXiv:1306.4669 [hep-th]} \BibitemShut
{NoStop}%
\bibitem [{\citenamefont {Yanagida}(1979)}]{Yanagida:1979as}%
\BibitemOpen
\bibfield  {author} {\bibinfo {author} {\bibfnamefont {T.}~\bibnamefont
		{Yanagida}},\ }\bibfield  {booktitle} {\emph {\bibinfo {booktitle}
		{Proceedings: Workshop on the Unified Theories and the Baryon Number in the
			Universe: Tsukuba, Japan, February 13-14, 1979}},\ }\href@noop {} {\bibfield
	{journal} {\bibinfo  {journal} {Conf. Proc.}\ }\textbf {\bibinfo {volume}
		{C7902131}},\ \bibinfo {pages} {95} (\bibinfo {year} {1979})}\BibitemShut
{NoStop}%
\bibitem [{\citenamefont {Schechter}\ and\ \citenamefont
	{Valle}(1980)}]{Schechter:1980gr}%
\BibitemOpen
\bibfield  {author} {\bibinfo {author} {\bibfnamefont {J.}~\bibnamefont
		{Schechter}}\ and\ \bibinfo {author} {\bibfnamefont {J.~W.~F.}\ \bibnamefont
		{Valle}},\ }\href {\doibase 10.1103/PhysRevD.22.2227} {\bibfield  {journal}
	{\bibinfo  {journal} {Phys. Rev.}\ }\textbf {\bibinfo {volume} {D22}},\
	\bibinfo {pages} {2227} (\bibinfo {year} {1980})}\BibitemShut {NoStop}%
\bibitem [{\citenamefont {Mohapatra}\ and\ \citenamefont
	{Senjanovic}(1980)}]{mohapatra:1980ia}%
\BibitemOpen
\bibfield  {author} {\bibinfo {author} {\bibfnamefont {R.~N.}\ \bibnamefont
		{Mohapatra}}\ and\ \bibinfo {author} {\bibfnamefont {G.}~\bibnamefont
		{Senjanovic}},\ }\href@noop {} {\bibfield  {journal} {\bibinfo  {journal}
		{Phys. Rev. Lett.}\ }\textbf {\bibinfo {volume} {44}},\ \bibinfo {pages}
	{912} (\bibinfo {year} {1980})}\BibitemShut {NoStop}%
\bibitem [{\citenamefont {Vilenkin}(1981)}]{Vilenkin:1981zs}%
\BibitemOpen
\bibfield  {author} {\bibinfo {author} {\bibfnamefont {A.}~\bibnamefont
		{Vilenkin}},\ }\href {\doibase 10.1103/PhysRevD.23.852} {\bibfield  {journal}
	{\bibinfo  {journal} {Phys. Rev.}\ }\textbf {\bibinfo {volume} {D23}},\
	\bibinfo {pages} {852} (\bibinfo {year} {1981})}\BibitemShut {NoStop}%
\bibitem [{\citenamefont {Mohapatra}\ and\ \citenamefont
	{Valle}(1986)}]{Mohapatra:1986bd}%
\BibitemOpen
\bibfield  {author} {\bibinfo {author} {\bibfnamefont {R.~N.}\ \bibnamefont
		{Mohapatra}}\ and\ \bibinfo {author} {\bibfnamefont {J.~W.~F.}\ \bibnamefont
		{Valle}},\ }\href {\doibase 10.1103/PhysRevD.34.1642} {\bibfield  {journal}
	{\bibinfo  {journal} {Phys. Rev.}\ }\textbf {\bibinfo {volume} {D34}},\
	\bibinfo {pages} {1642} (\bibinfo {year} {1986})}\BibitemShut {NoStop}%
\bibitem [{\citenamefont {Gonzalez-Garcia}\ and\ \citenamefont
	{Valle}(1989)}]{GonzalezGarcia:1988rw}%
\BibitemOpen
\bibfield  {author} {\bibinfo {author} {\bibfnamefont {M.~C.}\ \bibnamefont
		{Gonzalez-Garcia}}\ and\ \bibinfo {author} {\bibfnamefont {J.~W.~F.}\
		\bibnamefont {Valle}},\ }\href {\doibase 10.1016/0370-2693(89)91131-3}
{\bibfield  {journal} {\bibinfo  {journal} {Phys. Lett.}\ }\textbf {\bibinfo
		{volume} {B216}},\ \bibinfo {pages} {360} (\bibinfo {year}
	{1989})}\BibitemShut {NoStop}%
\bibitem [{\citenamefont {Joshipura}\ and\ \citenamefont
	{Valle}(1993)}]{Joshipura:1992hp}%
\BibitemOpen
\bibfield  {author} {\bibinfo {author} {\bibfnamefont {A.~S.}\ \bibnamefont
		{Joshipura}}\ and\ \bibinfo {author} {\bibfnamefont {J.~W.~F.}\ \bibnamefont
		{Valle}},\ }\href {\doibase 10.1016/0550-3213(93)90337-O} {\bibfield
	{journal} {\bibinfo  {journal} {Nucl. Phys.}\ }\textbf {\bibinfo {volume}
		{B397}},\ \bibinfo {pages} {105} (\bibinfo {year} {1993})}\BibitemShut
{NoStop}%
\bibitem [{\citenamefont {Berezhiani}\ \emph {et~al.}(1992)\citenamefont
	{Berezhiani}, \citenamefont {Smirnov},\ and\ \citenamefont
	{Valle}}]{Berezhiani:1992cd}%
\BibitemOpen
\bibfield  {author} {\bibinfo {author} {\bibfnamefont {Z.~G.}\ \bibnamefont
		{Berezhiani}}, \bibinfo {author} {\bibfnamefont {A.~{\relax Yu}.}\
		\bibnamefont {Smirnov}}, \ and\ \bibinfo {author} {\bibfnamefont {J.~W.~F.}\
		\bibnamefont {Valle}},\ }\href {\doibase 10.1016/0370-2693(92)90126-O}
{\bibfield  {journal} {\bibinfo  {journal} {Phys. Lett.}\ }\textbf {\bibinfo
		{volume} {B291}},\ \bibinfo {pages} {99} (\bibinfo {year} {1992})},\ \Eprint
{http://arxiv.org/abs/hep-ph/9207209} {arXiv:hep-ph/9207209 [hep-ph]}
\BibitemShut {NoStop}%
\bibitem [{\citenamefont {Weinberg}(1980)}]{weinberg:1980bf}%
\BibitemOpen
\bibfield  {author} {\bibinfo {author} {\bibfnamefont {S.}~\bibnamefont
		{Weinberg}},\ }\href@noop {} {\bibfield  {journal} {\bibinfo  {journal}
		{Phys. Rev.}\ }\textbf {\bibinfo {volume} {D22}},\ \bibinfo {pages} {1694}
	(\bibinfo {year} {1980})}\BibitemShut {NoStop}%
\bibitem [{\citenamefont {Dvali}\ \emph {et~al.}(1998)\citenamefont {Dvali},
	\citenamefont {Lazarides},\ and\ \citenamefont {Shafi}}]{Dvali:1997uq}%
\BibitemOpen
\bibfield  {author} {\bibinfo {author} {\bibfnamefont {G.~R.}\ \bibnamefont
		{Dvali}}, \bibinfo {author} {\bibfnamefont {G.}~\bibnamefont {Lazarides}}, \
	and\ \bibinfo {author} {\bibfnamefont {Q.}~\bibnamefont {Shafi}},\ }\href
{\doibase 10.1016/S0370-2693(98)00145-2} {\bibfield  {journal} {\bibinfo
		{journal} {Phys. Lett.}\ }\textbf {\bibinfo {volume} {B424}},\ \bibinfo
	{pages} {259} (\bibinfo {year} {1998})},\ \Eprint
{http://arxiv.org/abs/hep-ph/9710314} {arXiv:hep-ph/9710314 [hep-ph]}
\BibitemShut {NoStop}%
\bibitem [{\citenamefont {Berezhiani}(1983)}]{Berezhiani:1983hm}%
\BibitemOpen
\bibfield  {author} {\bibinfo {author} {\bibfnamefont {Z.~G.}\ \bibnamefont
		{Berezhiani}},\ }\href {\doibase 10.1016/0370-2693(83)90737-2} {\bibfield
	{journal} {\bibinfo  {journal} {Phys. Lett.}\ }\textbf {\bibinfo {volume}
		{129B}},\ \bibinfo {pages} {99} (\bibinfo {year} {1983})}\BibitemShut
{NoStop}%
\bibitem [{\citenamefont {Berezhiani}\ and\ \citenamefont
	{Khlopov}(1990)}]{Berezhiani:1990wn}%
\BibitemOpen
\bibfield  {author} {\bibinfo {author} {\bibfnamefont {Z.~G.}\ \bibnamefont
		{Berezhiani}}\ and\ \bibinfo {author} {\bibfnamefont {M.~{\relax Yu}.}\
		\bibnamefont {Khlopov}},\ }\href@noop {} {\bibfield  {journal} {\bibinfo
		{journal} {Sov. J. Nucl. Phys.}\ }\textbf {\bibinfo {volume} {51}},\ \bibinfo
	{pages} {739} (\bibinfo {year} {1990})},\ \bibinfo {note} {[Yad.
	Fiz.51,1157(1990)]}\BibitemShut {NoStop}%
\bibitem [{\citenamefont {Lazarides}\ and\ \citenamefont
	{Shafi}(1982)}]{Lazarides:1982tw}%
\BibitemOpen
\bibfield  {author} {\bibinfo {author} {\bibfnamefont {G.}~\bibnamefont
		{Lazarides}}\ and\ \bibinfo {author} {\bibfnamefont {Q.}~\bibnamefont
		{Shafi}},\ }\href {\doibase 10.1016/0370-2693(82)90506-8} {\bibfield
	{journal} {\bibinfo  {journal} {Phys. Lett.}\ }\textbf {\bibinfo {volume}
		{115B}},\ \bibinfo {pages} {21} (\bibinfo {year} {1982})}\BibitemShut
{NoStop}%
\bibitem [{\citenamefont {Barr}\ \emph {et~al.}(1982)\citenamefont {Barr},
	\citenamefont {Reiss},\ and\ \citenamefont {Zee}}]{Barr:1982bb}%
\BibitemOpen
\bibfield  {author} {\bibinfo {author} {\bibfnamefont {S.~M.}\ \bibnamefont
		{Barr}}, \bibinfo {author} {\bibfnamefont {D.~B.}\ \bibnamefont {Reiss}}, \
	and\ \bibinfo {author} {\bibfnamefont {A.}~\bibnamefont {Zee}},\ }\href
{\doibase 10.1016/0370-2693(82)90331-8} {\bibfield  {journal} {\bibinfo
		{journal} {Phys. Lett.}\ }\textbf {\bibinfo {volume} {116B}},\ \bibinfo
	{pages} {227} (\bibinfo {year} {1982})}\BibitemShut {NoStop}%
\bibitem [{\citenamefont {Bonilla}\ and\ \citenamefont
	{Valle}(2016)}]{Bonilla:2016zef}%
\BibitemOpen
\bibfield  {author} {\bibinfo {author} {\bibfnamefont {C.}~\bibnamefont
		{Bonilla}}\ and\ \bibinfo {author} {\bibfnamefont {J.~W.~F.}\ \bibnamefont
		{Valle}},\ }\href {\doibase 10.1016/j.physletb.2016.09.022} {\bibfield
	{journal} {\bibinfo  {journal} {Phys. Lett.}\ }\textbf {\bibinfo {volume}
		{B762}},\ \bibinfo {pages} {162} (\bibinfo {year} {2016})},\ \Eprint
{http://arxiv.org/abs/1605.08362} {arXiv:1605.08362 [hep-ph]} \BibitemShut
{NoStop}%
\bibitem [{\citenamefont {Bonilla}\ \emph {et~al.}(2016)\citenamefont
	{Bonilla}, \citenamefont {Ma}, \citenamefont {Peinado},\ and\ \citenamefont
	{Valle}}]{Bonilla:2016diq}%
\BibitemOpen
\bibfield  {author} {\bibinfo {author} {\bibfnamefont {C.}~\bibnamefont
		{Bonilla}}, \bibinfo {author} {\bibfnamefont {E.}~\bibnamefont {Ma}},
	\bibinfo {author} {\bibfnamefont {E.}~\bibnamefont {Peinado}}, \ and\
	\bibinfo {author} {\bibfnamefont {J.~W.~F.}\ \bibnamefont {Valle}},\ }\href
{\doibase 10.1016/j.physletb.2016.09.027} {\bibfield  {journal} {\bibinfo
		{journal} {Phys. Lett.}\ }\textbf {\bibinfo {volume} {B762}},\ \bibinfo
	{pages} {214} (\bibinfo {year} {2016})},\ \Eprint
{http://arxiv.org/abs/1607.03931} {arXiv:1607.03931 [hep-ph]} \BibitemShut
{NoStop}%
\bibitem [{\citenamefont {Bento}\ and\ \citenamefont
	{Valle}(1991)}]{Bento:1991bc}%
\BibitemOpen
\bibfield  {author} {\bibinfo {author} {\bibfnamefont {L.}~\bibnamefont
		{Bento}}\ and\ \bibinfo {author} {\bibfnamefont {J.~W.~F.}\ \bibnamefont
		{Valle}},\ }\href {\doibase 10.1016/0370-2693(91)90364-V} {\bibfield
	{journal} {\bibinfo  {journal} {Phys.Lett.}\ }\textbf {\bibinfo {volume}
		{B264}},\ \bibinfo {pages} {373} (\bibinfo {year} {1991})}\BibitemShut
{NoStop}%
\bibitem [{\citenamefont {Peltoniemi}\ \emph {et~al.}(1993)\citenamefont
	{Peltoniemi}, \citenamefont {Tommasini},\ and\ \citenamefont
	{Valle}}]{Peltoniemi:1992ss}%
\BibitemOpen
\bibfield  {author} {\bibinfo {author} {\bibfnamefont {J.}~\bibnamefont
		{Peltoniemi}}, \bibinfo {author} {\bibfnamefont {D.}~\bibnamefont
		{Tommasini}}, \ and\ \bibinfo {author} {\bibfnamefont {J.~W.~F.}\
		\bibnamefont {Valle}},\ }\href {\doibase 10.1016/0370-2693(93)91837-D}
{\bibfield  {journal} {\bibinfo  {journal} {Phys.Lett.}\ }\textbf {\bibinfo
		{volume} {B298}},\ \bibinfo {pages} {383} (\bibinfo {year}
	{1993})}\BibitemShut {NoStop}%
\bibitem [{\citenamefont {Ma}\ and\ \citenamefont
	{Srivastava}(2015)}]{Ma:2014qra}%
\BibitemOpen
\bibfield  {author} {\bibinfo {author} {\bibfnamefont {E.}~\bibnamefont
		{Ma}}\ and\ \bibinfo {author} {\bibfnamefont {R.}~\bibnamefont
		{Srivastava}},\ }\href {\doibase 10.1016/j.physletb.2014.12.049} {\bibfield
	{journal} {\bibinfo  {journal} {Phys. Lett.}\ }\textbf {\bibinfo {volume}
		{B741}},\ \bibinfo {pages} {217} (\bibinfo {year} {2015})},\ \Eprint
{http://arxiv.org/abs/1411.5042} {arXiv:1411.5042 [hep-ph]} \BibitemShut
{NoStop}%
\bibitem [{\citenamefont {Ma}\ \emph {et~al.}(2015)\citenamefont {Ma},
	\citenamefont {Pollard}, \citenamefont {Srivastava},\ and\ \citenamefont
	{Zakeri}}]{Ma:2015mjd}%
\BibitemOpen
\bibfield  {author} {\bibinfo {author} {\bibfnamefont {E.}~\bibnamefont
		{Ma}}, \bibinfo {author} {\bibfnamefont {N.}~\bibnamefont {Pollard}},
	\bibinfo {author} {\bibfnamefont {R.}~\bibnamefont {Srivastava}}, \ and\
	\bibinfo {author} {\bibfnamefont {M.}~\bibnamefont {Zakeri}},\ }\href
{\doibase 10.1016/j.physletb.2015.09.010} {\bibfield  {journal} {\bibinfo
		{journal} {Phys. Lett.}\ }\textbf {\bibinfo {volume} {B750}},\ \bibinfo
	{pages} {135} (\bibinfo {year} {2015})},\ \Eprint
{http://arxiv.org/abs/1507.03943} {arXiv:1507.03943 [hep-ph]} \BibitemShut
{NoStop}%
\end{thebibliography}
%

\end{document}